\documentclass[journal=jctcce,manuscript=article]{achemso}
\setkeys{acs}{articletitle = true}

\usepackage[version=3]{mhchem} 
\usepackage{bm}
\usepackage{mathtools}
\usepackage{amssymb}
\usepackage{changes}

\newcommand{\mbeq}{\overset{!}{=}}

\author{Bj{\"o}rn Kirchhoff}
\affiliation[HI]
{Science Institute and Faculty of Physical Sciences, University of Iceland, VR-III, 107 Reykjavík (Iceland).}
\author{Elvar Örn Jónsson}
\affiliation[HI]
{Science Institute and Faculty of Physical Sciences, University of Iceland, VR-III, 107 Reykjavík (Iceland).}
\author{Asmus Ougaard Dohn}
\affiliation[HI]
{Science Institute and Faculty of Physical Sciences, University of Iceland, VR-III, 107 Reykjavík (Iceland).}
\alsoaffiliation[DTU]
{Technical University of Denmark, Lyngby, Denmark.}
\author{Timo Jacob}
\affiliation[UUlm]
{Institute of Electrochemistry, Ulm University, Albert-Einstein-Allee 47, 89081 Ulm (Germany).}
\alsoaffiliation[HIU]
{Helmholtz-Institute Ulm (HIU) Electrochemical Energy Storage, Helmholtz-Straße 16, 89081 Ulm (Germany).}
\alsoaffiliation[KIT]
{Karlsruhe Institute of Technology (KIT), P.O. Box 3640, 76021 Karlsruhe (Germany).}
\author{Hannes J{\'o}nsson}
\email{hj@hi.is}
\affiliation[HI]
{Science Institute and Faculty of Physical Sciences, University of Iceland, VR-III, 107 Reykjavík (Iceland).}

\title{Elastic Collision Based Dynamic Partitioning Scheme for Hybrid Simulations}


\begin{document}







\begin{abstract}

 The scattering-adapted flexible inner region ensemble separator (SAFIRES) is a partitioning scheme designed to divide a simulation cell into two regions to be treated with different computational methodologies. SAFIRES prevents particles from crossing between regions and resolves boundary events through elastic collisions of the particles mediated by the boundary, conserving energy and momenta. A multiple-time-step propagation algorithm is introduced where the time step is scaled automatically to identify the moment a collision occurs. If the length of the time step is kept constant, the new propagator reduces to a regular algorithm for Langevin dynamics, and to the velocity Verlet algorithm for classical dynamics if the friction coefficient is set to zero. SAFIRES constitutes the exact limit of the premise behind boundary-based methods such as FIRES, BEST, and BCC which take advantage of the indistinguishability of molecules on opposite sides of the separator. It gives correct average ensemble statistics despite the introduction of an ensemble separator. SAFIRES is tested in simulations where the molecules on the two sides are treated in the same way, for a Lennard-Jones (LJ) liquid and a LJ liquid in contact with a surface, as well as for liquid modelling simulations using the TIP4P force field. Simulations using SAFIRES are shown to reproduce the unconstrained reference simulations without significant deviations.
\end{abstract}


\section{Introduction}

With the advent of and widespread access to high-performance computing resources over the last three decades, computational methods have become increasingly important in predicting material properties and understanding chemical reaction mechanisms. However, the desire to correctly describe processes at interfaces has been straining the limits of standard computational methods based on density functional theory (DFT). The solid-liquid interface is of particular interest and has proven especially troublesome since solvents --- and especially water --- often take part in reactions rather than acting as neutral bystanders\cite{kitanosono_catalytic_2018}. The influence of solvation can alter reaction behavior significantly\cite{gould2020,mellmer2018,roman2006,rossin2006} and can even cause structural changes in the catalyst through solvent-induced surface rearrangement\cite{warzok2018,yang2019}. Correct description of solvent interactions is oftentimes crucial when trying to understand complex natural phenomena such as the water splitting reaction in photosystem II\cite{hodel2016} or substrate binding in protein active sites\cite{abel2008}. 

Unfortunately, only few explicit solvent molecules can be included in DFT calculations due to quickly inflating computational effort. Furthermore, static DFT calculations, even when human bias is limited via the use of global optimization methods\cite{burnham2019,zhang2016,reda2018}, may only serve as an approximation for the first rigid ice-like layers of \ce{H2O} on transition metal surfaces or for ordered water clusters. However, this approach is less useful for less strongly interacting molecules and for the description of bulk-like properties of the solvent where less ordered structures are expected.

When studying the electronic interactions of, for example, a solid-liquid interface system, the interactions that are most important for the correct description of the reactive properties of a model are those between the surface, the adsorbate, and the first few layers of solvent. Relevant interactions include charge transfer, hybridization of electronic states, and proton transfer reactions. The widely used implicit solvation models, which represent the solvent as a potential based on key bulk properties such as the dielectric constant and dipole moment, in combination with DFT calculations fail to capture these important effects and are often reported to not improve results over calculations without implicit solvation\cite{heenen2020,eskyner2015,gray2017,zhang2017}. 

One possible solution to this problem is to divide the simulation into two coupled regions which are treated using different computational methodologies. A prominent example of this approach are QM/MM\cite{warshel1976,thole1980,field1990} methods. Here, the surface, the active species, and few solvent molecules are calculated using accurate and expensive QM methods while the majority of the (entirely explicit) solvent molecules is computed using efficient force field methods. This significant computational speedup allows users to run molecular dynamics (MD) or Monte Carlo (MC) methods to sample the phase space and to extract the solvation energy as a thermodynamic average.

A crucial facet in any coupling scheme is a sound description of the boundary between inner and outer region. Several different approaches exist to implement this necessary but inherently nonphysical concept while causing the least amount of disturbance to the system. In adaptive methods, particles are allowed to exchange between inner and outer regions\cite{adptv1,adptv2,adptv3,adptv4,adptv5,adptv6,adptv7,adptv8,adptv9}. When different potentials are used in the description of the inner and outer region, this approach can lead to discontinuities where the potentials come into contact\cite{adptv9}. In order to retain smooth dynamics and correct ensemble properties when particles cross between regions, a large number of different configurations of particles in a buffer region around the boundary have to be calculated if consistency of the total energy and forces is to be achieved. This corrective process increases computational costs by orders of magnitude.

Another boundary approach is presented by Rowley and Roux in form of the Flexible Inner Region Ensemble Separator (FIRES)\cite{rowley2012}. With FIRES, particles cannot transfer between regions but the boundary can expand and contract based on the position of the outermost particle in the inner region. This implies that a typical FIRES simulation contains three distinct parts: a solute that acts as the central point of the simulation, an inner region of particles around the solute, and an outer region of particles; the solute and the inner region are treated using the same computational methodology. The boundary radius is defined as the distance between the solute and the outermost particle in the inner region and enforced using a Hookean force constraint acting on the inner and outer particles. 

While this method sacrifices microscopic trajectories of individual particles in the system, it can be mathematically shown to deliver correct average thermodynamic values\cite{beglov_finite_1994,rowley2012} if the same chemical species is present in both regions. FIRES therefore establishes the main premise that all boundary-based separation algorithms rely on: it is assumed that a situation where a particle from the outer region travels to the inside \textit{et vice verse} is statistically equivalent to a situation where the particles are redirected back where they came from at the point of equal distance to the solute. Thus, correct average thermodynamic properties can be retained this way at the cost of sacrificing continuity of individual particle trajectories. FIRES was used to good effect in benchmark calculations for DFTB3 paramterization of Zn and Mg\cite{lu2015} and to calculate the solvation of a transition state of a nucleophilic carbonyl reaction\cite{boereboom2018}.

Notably, the finite force constraint used in the Rowley and Roux implementation is actually an approximation to the FIRES premise, which in its mathematical formalism implies instantaneous redirection at the boundary. It is conceivable that this approximation could lead to what Bulo \textit{et al.} reported as an accumulation of molecules at the boundary when using FIRES, analogous to what is observed for coupled simulations with a static, hard sphere boundary\cite{bulo2013}.

Aside from FIRES, other boundary-based methods have been developed, in particular the Boundary based on Exchange Symmetry (BEST) method \cite{shiga2013,shiga_erratum_2013} and the Boundary Constraint with Correction (BCC) method\cite{takahashi2018}. BEST aims to find a generalized approach to the separating potential. The BEST separating potential is constructed by assigning particles to the inner or outer region and calculating a penalty function which is unity for a separated particle pair and approaches zero as the particle configuration becomes more undesirable, \textit{i.e.} as a particle pair travels further into the respective other region\cite{shiga2013,shiga_erratum_2013}. The bias function is then obtained as the product of penalty functions of all permutations of particle pairs; hence, the reference to exchange symmetry. However, since particle pairs that are unproblematic (\textit{i.e.} in different phases) do not contribute significantly to the bias function, Shiga and Masia find that construction of the bias function can be simplified by only taking into account single or double exchanges of the particle pairs with the penalty functions closest to zero.

The BCC method utilizes Fermi-derived bias potential inspired by BEST to separate particles in the regions\cite{takahashi2018}. As the authors remark, the basic premise behind boundary-based methods --- instantaneous redirection at the boundary --- cannot be achieved using a constraining potential. The BCC method therefore attempts to rectify this statistical error, which is related to the accumulation artifacts at the boundary discussed for FIRES, by performing an additional set of calculations to remove contributions of the bias potential from thermodynamic properties of the erroneous configurations. This approach therefore adds additional computational overhead to simulations.

\begin{figure}[htbp]
    \centering
        \includegraphics[width=\linewidth]{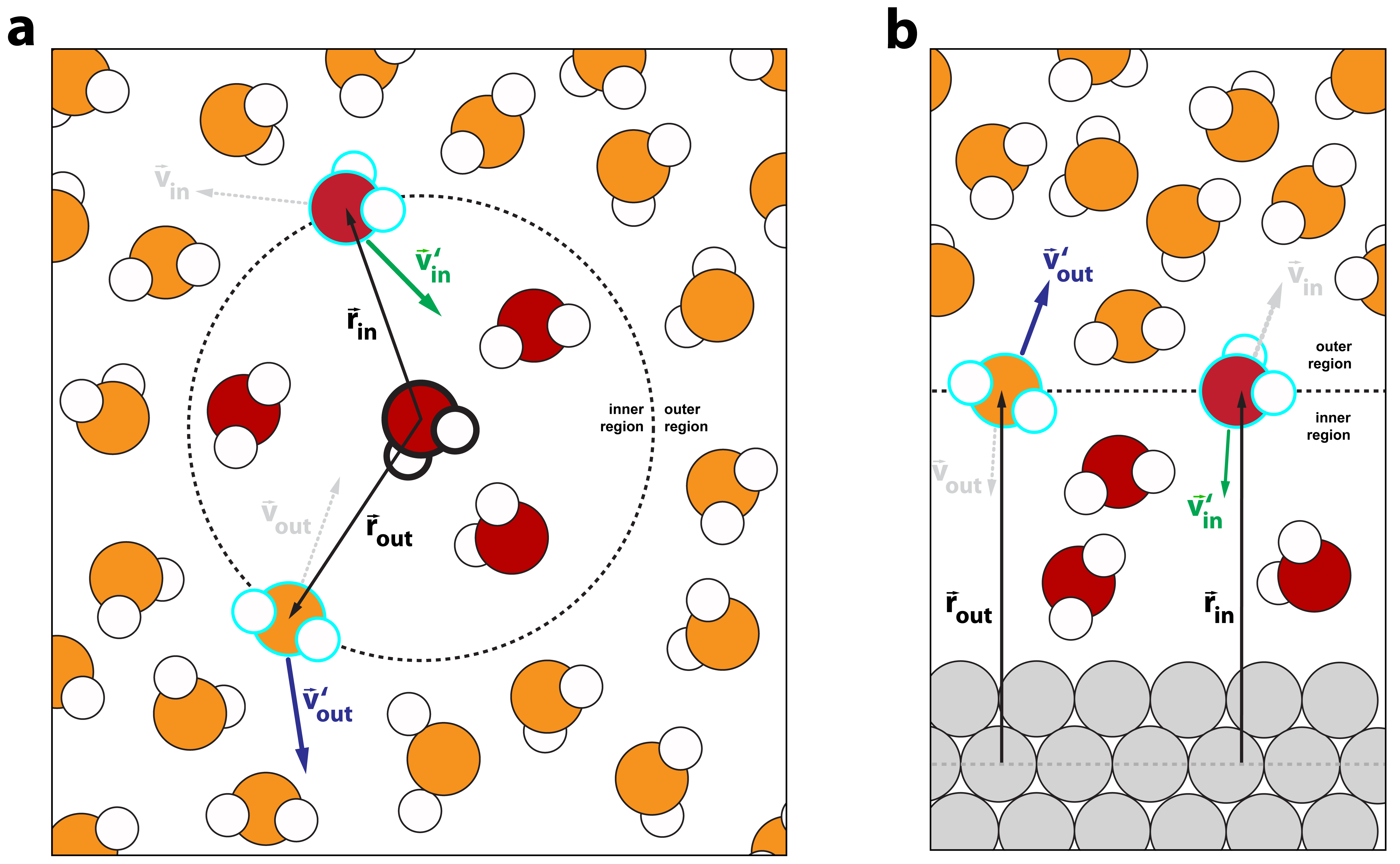}
    \caption{Illustration of the SAFIRES approach for a molecular (\textbf{a}) and an infinite-surface based model system (\textbf{b}) with water. The flexible boundary is shown as a dotted black circle or line. Molecules in the inner region are shown in red, those in the outer region are shown in orange. The surface is shown in grey. $r$: vector connecting the solute (\textbf{a}) or a plane parallel to the surface (\textbf{b}) to the particles in the inner (in) and outer (out) region. The greyed out and dotted $\vec{v}$ vectors represent the initial velocity of the molecule in the inner (in) and outer (out) region before the SAFIRES algorithm redirects them. The vectors $\vec{v}'$ represent the velocity vectors of the molecule in the inner (in, green) and outer (out, blue) region after the SAFIRES algorithm redirected them via an elastic collision mediated by the boundary.}
    \label{fig:flashy_illustration}
\end{figure}

In this study, a new separation algorithm, the scattering-adapted flexible inner region ensemble separator (SAFIRES), is introduced. SAFIRES divides a computational model into two regions to treat them using different methodologies. Analogous to other boundary-based methods, the boundary between the inner and outer region is defined flexibly between the solute (which can be molecular or periodic) and the outermost particle in the inner region. To enforce the boundary, SAFIRES performs elastic collisions between particles from the inner and outer region when they are at the same distance from the solute. SAFIRES therefore fully conserves energy and momenta. This instantaneous approach to particle redirection however necessitates that time steps be flexible during the simulation for collisions to be exact. Therefore, fractional time steps are calculated on-the-fly as conflicts arise and a new multiple-time-step propagation algorithm is introduced. The new propagator reduces to the Vanden-Eijnden / Ciccotti implementation of the Langevin algorithm\cite{vanden-eijnden_second-order_2006} for constant time steps and further reduces to the Velocity-Verlet algorithm for constant time steps and a friction coefficient of zero. As a final requirement, forces are only updated at full intervals of the initial time step in order to retain the temporal consistency of the calculation and to satisfy time-reversibility of the Taylor expressions that the propagation algorithm is derived from. Figure \ref{fig:flashy_illustration} illustrates the SAFIRES approach for molecular and surface model systems in a schematic way. 

This study is structured as follows: section \ref{s:safires_method} details the implementation of SAFIRES. Therein, subsection \ref{s:overview} gives an overview of the SAFIRES algorithm, subsection \ref{s:timestep} details propagation and how fractional time steps are calculated to find the exact point of collision, subsection \ref{s:collision} examines the collision handling, and subsection \ref{s:advanced} delves into advanced details of the implementation. Subsequently, sections \ref{s:computational} and \ref{s:test} describe the computational details and test calculations of SAFIRES, respectively. The SAFIRES algorithm is first illustrated based on a Lennard-Jones (LJ) liquid model system with a molecular solute (subsection \ref{s:lj_liquid}) before expanding the approach to a LJ infinite-surface model as the solute for applications at the solid-liquid interface (subsection \ref{s:lj_surface}). Inert noble gas surfaces have been used in the past, for example in a QM/MM study by Daru \textit{et al.} in order to study the rearrangement of ion-doped water clusters on an inert surface\cite{daru2019}. Partial RDFs can be used to study the structure of a solvent near a surface\cite{agrafonov2015}. Finally, SAFIRES is tested in an MM*/MM simulation of a water-in-water model using the TIP4P force field (subsection \ref{s:tip4p}). Radial distribution functions (RDFs) are used in all cases to evaluate the ability of SAFIRES to reproduce the correct average structural features of unconstrained simulations.

\section{The SAFIRES Method} \label{s:safires_method}

\subsection{Algorithm overview} \label{s:overview}

\begin{figure}[h!]
    \centering
    \includegraphics[width=\linewidth]{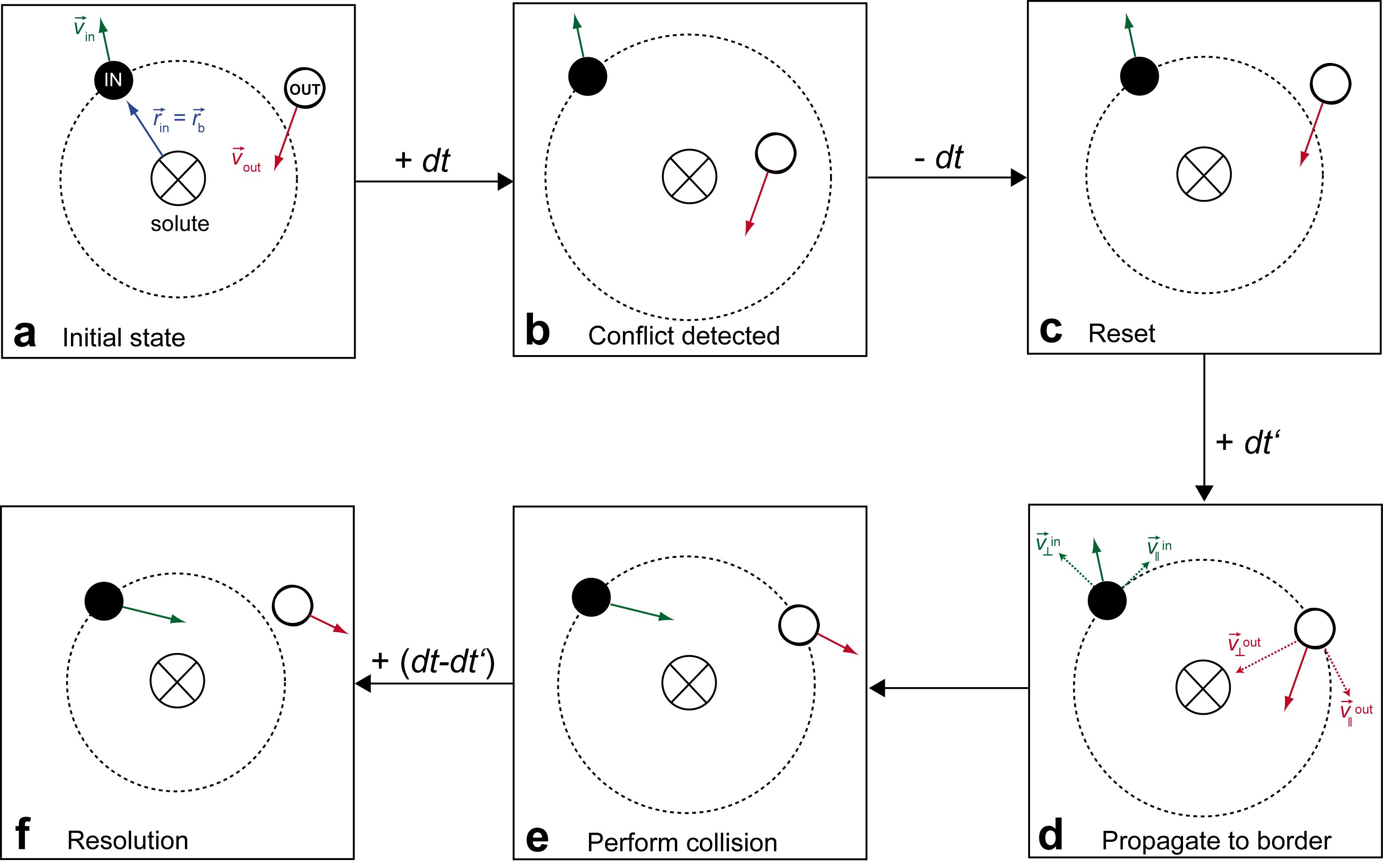}
    \caption{Schematic overview of the steps involved in the SAFIRES scheme. \textbf{a} The initial, conflict-free state of the system. $\Vec{r}_\text{b}$ indicates the boundary radius, which is defined by the distance between the solute and the outermost inner region (in) particle. $\Vec{v}_\text{in}$ and $\Vec{v}_\text{out}$ indicate the velocity vectors of an inner and outer region particle, respectively. \textbf{b} After the system is propagated using the default time step $dt$, the SAFIRES algorithm checks if one or more outer particles are closer to the solute than the outermost inner particle. If a conflict is detected, the SAFIRES algorithm  proceeds to resolve the issue. \textbf{c} The first step to resolve the issue is to reset the configuration to the last conflict-free state. \textbf{d} A fractional time step $dt'$ is solved for and executed which exactly propagates the conflicting in/out particle pair to the boundary. $\Vec{v}_\perp$ and $\Vec{v}_\parallel$ indicate the normal and tangential velocity components, respectively. \textbf{e} An elastic collision between the conflicting in/out pair is performed,
    exchanging propagating components normal to the boundary surface. \textbf{f} After the collision has changed the trajectories of the conflicting in/out pair, the system is propagated by the remaining time step, $dt'-dt$, required to fulfill a complete regular time step. Afterwards, SAFIRES checks if another outer particle has propagated beyond the outermost inner particle. If this is the case, SAFIRES repeats steps \textbf{c}-\textbf{f} until there no more conflicts detected and a full
    time step $dt$ is achieved.
    }
    \label{fig:safires_scheme}
\end{figure}

The key steps in SAFIRES for detecting and resolving a crossing of particles from the outer to the inner region are illustrated with a spherically-symmetric model system, as depicted schematically in Figure \ref{fig:safires_scheme}. 
For simplicity it is assumed that all particles are monoatomic and that the solute particle is fixed; particles with more than one atom are discussed in section \ref{s:advanced}. The SAFIRES routine 
for a 2D-periodic system is largely the same with some simplifications as outlined in
Section~\ref{s:2D-pbc}.

At the start of an MD simulation, three sets of indices are assigned. The first, $s$, identifies the solute particle, whose center of mass acts as the origin for the boundary radius. The second set, $S_\text{in}$, identifies the particles within this radius, and lastly, $S_\text{out}$ identifies the particles outside of this boundary. 

Given the 3$n$-dimensional position vector field of all $n$ particles of the system, $\mathbf{R}^i$, at iteration $i=0$, the index assignment has to satisfy 
\begin{equation}
    \{\mid\mid \mathbf{R}^{i}(\alpha) - \mathbf{R}^i(s)\mid\mid\ \forall \alpha \in S_\text{in} \} <\  
    \{\mid\mid \mathbf{R}^{i}(\beta) - \mathbf{R}^i(s)\mid\mid\ \forall \beta \in S_\text{out}\}\,, \label{eq:cond1}
\end{equation}
i.e. all of the outer particles should be further away from
the solute compared to all of the inner
particles.  $||\mathbf{R}||$ denotes the Euclidian norm. 
Note that the condition simply involves a change of origin of the input coordinate space, and any fixed point of origin can in principle be used -- it does not have to be associated with a particle.


The SAFIRES algorithm is called after an integrated time step $dt>0$ of the superordinate MD simulation is performed which propagates the particle positions, leading to
\begin{equation}
    \mathbf{R}^i \xrightarrow{dt} \mathbf{R}^{i+1}\,.
\end{equation}
First, the radius of the boundary sphere, $r_b$, which is shown as a dotted circle in Figure \ref{fig:safires_scheme}, is updated. The radius is based on the farthest distance between the fixed solute (s) and inner (in) particles
\begin{equation}
    \alpha_b \leftarrow r_b = \text{max}\left\{\mid\mid \mathbf{R}^{i+1}(\alpha) - \mathbf{R}^{i+1}(s)\mid\mid \forall \alpha \in S_\text{in}\right\}
    \label{eq:boundary}
\end{equation}
Here $\alpha_b$ indexes the outermost inner particle.
SAFIRES then checks if any of the outer (out) particles have crossed the boundary
\begin{equation}
    S'_\text{out} = \left\{\mid\mid \mathbf{R}^{i+1}(\beta) - \mathbf{R}^{i+1}(s)\mid\mid < r_b\ \forall \beta \in S_\text{out}\right\}
    \label{eq:conflict}
\end{equation}
where $S'_\text{out}$ is a subset of $S_\text{out}$ and identifies the conflicting outer particles (it can have more than one element).
If no conflicts are detected (i.e. $S'_\text{out} = \O$), the systems state is stored 
and the next iteration of the MD integrator performed. 

If a conflict is detected (Figure \ref{fig:safires_scheme}\textbf{b}), the simulation is reverted to the last conflict-free system state (Figure \ref{fig:safires_scheme}\textbf{c}). 
From this state, a fractional time step $dt' \leq dt$ is determined -- for all conflicting particle pairs as indexed by $\alpha_b$ and $S'_\text{out}$ -- and the smallest $dt'$ executed in order to propagate the conflicting in/out particle pair to the same distance from the solute 
\begin{equation}
    \mathbf{R}^{i} \xrightarrow{dt'} \mathbf{R}^{i+\delta}\,,
\end{equation}
where $\delta$ indicates that only a partial iteration is performed (which can be defined as $\delta = dt'/dt$ for completeness). The smallest value of $dt'$ is resolved first since it 
corresponds to the first boundary crossing in the time step interval $dt$.
The time step extrapolation that yields $dt'$ is discussed in section \ref{s:timestep}. 

The partial propagation, leading to the state shown in Figure \ref{fig:safires_scheme}\textbf{d}, is handled by SAFIRES.
In this state, an elastic collision is performed between the conflicting in/out particle pair resulting in an exchange of propagating vectors normal to the boundary surface, $\mathbf{v}_\perp^b$, resulting in Figure \ref{fig:safires_scheme}\textbf{e}.
This step is detailed in section \ref{s:collision}. After the collision has been performed successfully, the resulting system state is saved in case the ensemble needs to be reverted again.

Lastly, the system is propagated by the remaining fractional time step $dt-dt'$ that is required to complete a regular time step $dt$, see Figure \ref{fig:safires_scheme}\textbf{f}. The propagation is performed by the SAFIRES propagator and uses the post-collision velocity vectors 
\begin{equation}
    \mathbf{R}^{i+\delta} \xrightarrow{(dt-dt'),\mathbf{v}_\perp^b} \mathbf{R}^{i+1}_\text{c}
\end{equation}
After this second fractional propagation step, the boundary is updated based on the new positions and another check for conflicts is performed using $\mathbf{R}^{i+1}_c$ in equations~\eqref{eq:boundary}-\eqref{eq:conflict}. This updates both the index $\alpha_b$ and index set $S'_\text{out}$ since having resolved the first boundary crossing any other inner and outer particle pair can become a new conflicting pair. If a new conflict arises the steps depicted in Figure \ref{fig:safires_scheme}{\bf c}-{\bf f} are repeated. 
Multiple conflicts are detailed in section \ref{s:advanced}.

After a conflict free system state is reached, the SAFIRES propagator completes the iteration. Control is then returned to the superordinate MD propagator and a new iteration is started.



\subsection{Time step extrapolation} \label{s:timestep}

The relationship for extrapolation of the time step required to propagate the conflicting in/out particle pair to the same distance from the solute is obtained from the Langevin propagator\cite{vanden-eijnden_second-order_2006}
\begin{eqnarray}
    \mathbf{V}^{i + \frac{1}{2}} &=& \mathbf{V}^i + \frac{1}{2} dt\mathbf{F}(\mathbf{R}^i) - \frac{1}{2} dt\gamma \mathbf{V}^i + \frac{1}{2} \sqrt{dt} \mathbf{\Sigma}^i \bm{\sigma} \nonumber \\
    &-& \frac{1}{8} dt^2 \gamma \left(\mathbf{F}(\mathbf{R}^i) - \gamma \mathbf{V}^i\right) - \frac{1}{4} dt^\frac{3}{2} \gamma \left( \frac{1}{2} \mathbf{\Sigma}^i + \frac{1}{\sqrt{3}} \mathbf{\Lambda}^i \right) \bm{\sigma} \label{eq:v_first_halfstep}\\
    \mathbf{R}^{i+1} &=& \mathbf{R}^i + dt \mathbf{V}^{i+\frac{1}{2}} + dt ^\frac{3}{2}  \frac{1}{2 \sqrt{3}} \mathbf{\Lambda}^i \bm{\sigma} \label{eq:x_step}\\
    \mathbf{V}^{i + 1} &=& \mathbf{V}^{i + \frac{1}{2}} + \frac{1}{2} dt \mathbf{F}(\mathbf{R}^{i+1}) - \frac{1}{2} dt \gamma \mathbf{V}^{i + \frac{1}{2}} + \frac{1}{2} \sqrt{dt} \mathbf{\Sigma}^i\bm{\sigma} \nonumber \\
    &-& \frac{1}{8} dt^2 \gamma \left(\mathbf{F}(\mathbf{R}^{i+1}) - \gamma \mathbf{V}^{i+\frac{1}{2}}\right) - \frac{1}{4} dt^\frac{3}{2} \gamma \left( \frac{1}{2} \mathbf{\Sigma}^i + \frac{1}{\sqrt{3}} \mathbf{\Lambda}^i \right)\bm{\sigma} \label{eq:v_second_halfstep}\\
    \bm{\sigma} &=& \sqrt{\frac{2 T \gamma}{\mathbf{M}}}  \label{eq:sigma}
\end{eqnarray}
Here, $\mathbf{V}$
and $\mathbf{F}$ are the 3$n$-dimensional velocity
and force vector fields, respectively, for all $n$ particles. Similarly and for the following sections, lowercase boldfaced letters -- for example $\mathbf{v}$ and $\mathbf{r}$ -- 
refer to their 3-dimensional single particle analogs.
$\gamma$ is the thermostat friction coefficient, and $\mathbf{\Sigma}$ and $\mathbf{\Lambda}$ are 3$n$-dimensional covariant Gaussian distributions that are randomized at the start of each iteration $i$. $T$ is the target temperature for the thermostat and $\mathbf{M}$ is a $n$-dimensional column vector of the particle masses. Setting $\gamma = 0$ reduces this set of equations to the Velocity-Verlet propagator. 

The propagation during the SAFIRES routine is split into two parts --- first to propagate towards the boundary, and then to propagate away from it again. Forces are only evaluated again after 
a full time step has been completed and the coordinate space is without additional conflicts. Therefore, equation \eqref{eq:v_second_halfstep} is performed after all conflicts are resolved. As a result, only
equations \eqref{eq:v_first_halfstep}-\eqref{eq:x_step} need to be considered for the time step extrapolation of conflicting in/out particle pairs.

In order to satisfy the time reversibility of the Taylor expressions that the Velocity Verlet and Langevin propagators are derived from, all components in equation \eqref{eq:v_first_halfstep} and the third term in equation \eqref{eq:x_step} are added based on the full time step, $dt$. 
Equation \eqref{eq:x_step} is then reduced to
\begin{eqnarray}
    \mathbf{R}^{i+1} &=& \mathbf{R}^i + dt\mathbf{A}^{i} 
    \label{eq:reduced_x}\\
    \mathbf{A}^i &=& \mathbf{V}^{i+\frac{1}{2}} + \mathbf{B}^i \label{eq:A} \\
    \mathbf{B}^i &=& dt ^\frac{1}{2} \frac{1}{2 \sqrt{3}} \mathbf{\Lambda}^i \bm{\sigma}\label{eq:b_explanation}
\end{eqnarray}
and the propagation of the coordinate space in equation \eqref{eq:reduced_x} is now linear in terms of $dt$.
Given $\mathbf{R}^{i+1}$ as defined above, the outermost inner particle is identified via equation 
\eqref{eq:boundary}, and the condition of equation \eqref{eq:conflict} is checked. For $\alpha_b$
and elements $\beta$ in $S'_\text{out}$ all conflicting in/out particle pairs are considered. 
For the time step extrapolation, the key quantities are, however, not the particle positions $\mathbf{R}^i(\alpha_b)$ and $\mathbf{R}^i(S'_\text{out})$ but the distances 
of the in/out particle pair from the solute, which are more naturally traced with a change in origin
$\mathbf{R}^i_s = \mathbf{R}^i - \mathbf{R}^i(s)$. Hence, in this frame the norm
of the positions of the conflicting in/out pairs must be equal at the point of collision:
\begin{eqnarray}
    ||\mathbf{R}^{i+\delta}_s(\alpha_b)|| &\mbeq& ||\mathbf{R}^{i+\delta}_s(\beta)|| \ \ \forall \beta \in S'_\text{out} \label{eq:rin_rout}.
\end{eqnarray}
Using equation \eqref{eq:rin_rout} in combination with equation \eqref{eq:reduced_x} leads to
\begin{equation}
    ||\mathbf{R}_s^i(\alpha_b) + \delta t'(\beta) \mathbf{A}^i(\alpha_b)|| = ||\mathbf{R}_s^{i}(\beta) 
    + \delta t'(\beta) \mathbf{A}^i(\beta)|| \ \ \forall \beta \in S'_\text{out}\label{eq:solve_this}
\end{equation}
and is solved analytically for $\delta t'(\beta)$, which are the partial time steps required to propagate each conflicting in/out particle pairs to the same distance from the solute. Using 
\begin{equation}
    \beta_b \leftarrow dt' = \text{min}\{\delta t'(\beta)\mid \forall \delta t'(\beta)\in \mathbb{R}_+\}\label{eq:dt_solution}
\end{equation}
the temporally first boundary conflict and corresponding outer particle index, $\beta_b$, is identified, \textit{i.e.} the smallest positive real valued solution.
The system's coordinate space is evolved
accordingly
\begin{equation}
    \mathbf{R}^{i+\delta} = \mathbf{R}^i + dt'\mathbf{A}^i
    \label{eq:partial}
\end{equation}
leading to the state depicted in Figure \ref{fig:safires_scheme}\textbf{d}. Note that a positive real valued solution where $dt'\leq dt$ 
is always guaranteed since $dt>0$ and due to 
the condition set by equations~\eqref{eq:cond1} and \eqref{eq:conflict}. 

\subsection{Collision handling} \label{s:collision}

Once a conflict has been detected and the system propagated such that the inner ($\alpha_b$)
and outer ($\beta_b$) particle
are both at the boundary, 
the particles need to be redirected in order to resolve the conflict. This is achieved with an exchange of momentum mediated by adjusting the
propagating vectors normal to the boundary surface. In the case of Langevin dynamics the vectors $\mathbf{a}^i_\text{in}$ = $\mathbf{A}^i(\alpha_b)$ and $\mathbf{a}^i_\text{out} = \mathbf{A}^i(\beta_b)$ are considered. These components reduce to $\mathbf{v}^{i+\frac{1}{2}}_\text{in}$ and $\mathbf{v}^{i+\frac{1}{2}}_\text{out}$ in Velocity-Verlet. In the following the MD step index $i$ is omitted for clarity.

In case of a solute with a spherical boundary, a transformation of the coordinates of one of the involved particles is required to bring them into the same frame of reference. Geometrically speaking, the propagating vector of one particle needs to be rotated on the surface of the boundary sphere to the position of the other particle. To this end, consider the two vectors $\mathbf{r}_\text{in,s} = \mathbf{R}(\alpha_b) - \mathbf{R}(s)$ and $\mathbf{r}_\text{out,s}=\mathbf{R}(\beta_b) - \mathbf{R}(s)$ connecting the solute and the in/out particle pair. The angle $\phi$ between these vectors is given by Vincenty's formula\cite{vincenty1975} as
\begin{equation}
    \phi = \text{arctan2}\left( ||\mathbf{r}_\text{in,s} \times \mathbf{r}_\text{out,s}||, (\mathbf{r}_\text{in,s} \cdot \mathbf{r}_\text{out,s}) \right).
\end{equation}
Assuming rotation of the outer particle the rotated propagating vector is given by
\begin{equation}
    \mathbf{a}^\text{rot}_\text{out} = X(\mathbf{c},\phi) \cdot \mathbf{a}_\text{out},
\end{equation}
with a rotational matrix, $X(\mathbf{c},\phi)$, of the Euler-Rodrigues form and the rotational axis 
\begin{equation}
    \mathbf{c} = \frac{\mathbf{r}_\text{in,s} \times \mathbf{r}_\text{out,s}}{||\mathbf{r}_\text{in,s} \times \mathbf{r}_\text{out,s}||}.
\end{equation}

After $\mathbf{a}_\text{out}$ is transformed into the same reference frame as $\mathbf{a}_\text{in}$, an elastic collision is performed between the two particles according to 
\begin{eqnarray}
    \mathbf{a}_\perp^b 
    &=& \frac{((\mathbf{a}^\text{rot}_\text{out} - \mathbf{a}_\text{in}) \cdot \mathbf{r}_\text{in,s})\ \mathbf{r}_\text{in,s}}{||\mathbf{r}_\text{in,s}||^2} \label{eq:collision1} \\
    \mathbf{a}_\text{out}'^{\text{\ rot}} &=& \mathbf{a}_\text{out}^\text{rot} - \frac{2\ m_\text{in}}{m_\text{in} + m_\text{out}}\mathbf{a}_\perp^b\\
    \mathbf{a}_\text{in}' &=& \mathbf{a}_\text{in} + \frac{2\ m_\text{out}}{m_\text{in} +
    m_\text{out}}
    \mathbf{a}_\perp^b\label{eq:collision3}
\end{eqnarray}
where $\mathbf{a}_\perp^b$ is normal to the boundary surface, since it is is parallel to $\mathbf{r}_\text{in,s}$ which defines the boundary radius, and connects the solute to the particles $\alpha_\text{b}$ and $\beta_\text{b}$ which coincide after the rotation. In the 
case of Velocity Verlet this component is reduced to $\mathbf{v}_\perp^b$.


After the collision, $\mathbf{a}_\text{out}'^{\text{\ rot}}$ is rotated back to the outer particle 
reference frame 
\begin{equation}
    \mathbf{a}_\text{out}' = X(\mathbf{c}, 2\pi - \phi) \cdot \mathbf{a}_\text{out}'^{\text{\ rot}}
\end{equation}
and the system's propagating vector field is updated such that 
\begin{equation}
    \{\mathbf{A}^i(\alpha_b) = \mathbf{a}'_\text{in}, \mathbf{A}^i(\beta_b) =  \mathbf{a}_\text{out}'\} \to \mathbf{A}^i_c
\end{equation}
which implies 
\begin{equation}
    \{\mathbf{V}^{i+\frac{1}{2}}(\alpha_b) = \mathbf{v}'_\text{in}, \mathbf{V}^{i+\frac{1}{2}}(\beta_b) =  \mathbf{v}_\text{out}'\} \to \mathbf{V}^{i+\frac{1}{2}}_c
\end{equation}
since the components of
equation~\eqref{eq:A} are additive and both propagate the coordinate space 
linearly in terms of $dt$, such that the exchange as outlined above can be applied 
separately on each component.

The system coordinate space is evolved by the remainder of the time step
\begin{equation}
    \mathbf{R}^{i+1}_c = \mathbf{R}^{i+\delta} + (dt - dt')\mathbf{A}^i_c
\end{equation}
At this stage the updated coordinate space is checked for additional
conflicts, first by updating the boundary radius, equation \eqref{eq:boundary}, 
resulting in an update of $\alpha_b$, followed by a check if any outer 
particles have crossed the new boundary, equation \eqref{eq:conflict}, 
resulting in an update of $S'_\text{out}$. 
Multiple conflicts are discussed in section \ref{s:multiple}.

If no additional conflicts are detected in the interval $(dt - dt')$, the SAFIRES propagator triggers a force calculation for the new configuration $\mathbf{F}(\mathbf{R}^{i+1})$ and the velocity vectors are updated to $\mathbf{V}^{i+1}$ according
to equation \ref{eq:v_second_halfstep}, applied to 
$\mathbf{V}_c^{i+\frac{1}{2}}$.
This completes the update of all vector fields.


\subsection{Special cases} \label{s:advanced}

\subsubsection{Multiple conflicts during the same time step}\label{s:multiple}

Multiple conflicts do not add any additional complexity, rather,
the same procedures are applied as outlined in the preceding section, using the $\mathbf{R}^{i+\delta}$ and $\mathbf{A}^i_c$ 
vector fields as a starting point.
A new partial time step, $dt''$, is solved for following equations \eqref{eq:rin_rout}--\eqref{eq:dt_solution}, and is 
$dt''\leq (dt - dt')$, where the norm of the vector field $\mathbf{R}^{i+\delta}_s = \mathbf{R}^{i+\delta} - \mathbf{R}^{i+\delta}(s)$ is now considered. Namely the condition
\begin{eqnarray}
    ||\mathbf{R}^{i+\delta+\epsilon}_s(\alpha_b)|| &\mbeq& ||\mathbf{R}^{i+\delta+\epsilon}_s(\beta)|| \ \ \forall\ \beta \in S''_\text{out}\;,
\end{eqnarray}
where $\epsilon$ is defined as $dt''/(dt-dt')$, is imposed and solved for using
\begin{equation}
    ||\mathbf{R}_s^{i+\delta}(\alpha_b) + \delta t''(\beta) \mathbf{A}^i_c(\alpha_b)|| = ||\mathbf{R}^{i+\delta}_s(\beta) 
    + \delta t''(\beta) \mathbf{A}^i_c(\beta)|| \ \ \forall\ \beta \in S_\text{out}
\end{equation}
and
\begin{equation}
    \beta_b \leftarrow dt'' = \text{min}\{\delta t''(\beta)\mid \forall\ \delta t''(\beta)\in \mathbb{R}_+\}
\end{equation}
where $dt''$ now brings the temporally first conflict, corresponding to particles $\alpha_b$ and $\beta_b$, to the boundary radius as
defined by the propagation of the vector field
\begin{equation}
    \mathbf{R}^{i+\delta+\epsilon} = \mathbf{R}^{i+\delta} + dt''\mathbf{A}_\text{c}^i
    \label{eq:partial}
\end{equation}
An elastic collision is again performed using the vectors 
$\mathbf{a}^i_\text{c,in}$ and $\mathbf{a}^i_\text{c,out}$ (i.e. $\mathbf{A}_\text{c}^i(\alpha_b)$ and 
$\mathbf{A}_\text{c}^i(\beta_b)$ respectively).

This process
of resolving multiple conflicts and solving for partial time steps in smaller and smaller remaining time intervals
can in principle be continued indefinitely. However, in
the practical examples presented in this work (see Section X) 
we find it is a rare event for conflicts to exceed one, even if
multiple possible outer particles are identified through equation~
\eqref{eq:conflict}. This occurs since the resolution
of the temporally first conflict often
results in the resolution of other 
possible conflicts within the time step interval due to the change in the trajectory
of $\alpha_\text{b}$.

\subsubsection{\label{s:2D-pbc} Semi-infinite Surfaces}

In the case of a 2D-periodic system the SAFIRES 'check for conflicts' and 'elastic collision' steps are simplified. 
Given an orthorhombic left-handed Cartesian coordinate system with origin
$O=(0,0,0)$ and periodicity along the $x$- and $y$-axis, an $xy$-plane boundary is defined as the 
numerically largest $z$ position of the inner particles. The condition imposed on the indexing
of the inner and outer particles in the initial position vector space is then as follows
\begin{equation}
    \{\mathbf{R}_z^{i}(\alpha)\ \forall \alpha \in S_\text{in}\} <\  
    \{\mathbf{R}_z^{i}(\beta)\ \forall \beta \in S_\text{out}\} \label{eq:cond2}
\end{equation}
such that all outer particles are above the $xy$-plane boundary. Here $\mathbf{R}_z^i(\alpha)$ is shorthand for $\mathbf{R}^i(\alpha)(0,0,z)$, \textit{i.e.} the \textit{z}-coordinate of particle $\alpha$. After an integrated time step
of the MD simulation the resulting particle positions are used to identify the outermost inner
particle
\begin{equation}
    \alpha_b \leftarrow r_z = \text{max}\left\{\mathbf{R}_z^{i+1}(\alpha)\ \forall \alpha \in S_\text{in}\right\}
    \label{eq:boundary2}
\end{equation}
and the check for conflicts is then simply
\begin{equation}
    S'_\text{out} = \left\{\mathbf{R}_z^{i+1}(\beta) < r_z\ \forall \beta \in S_\text{out}\right\}
    \label{eq:conflict2}
\end{equation}
The partial time step required to propagate the conflicting particles to the boundary is then
solved for according to
\begin{eqnarray}
    \mathbf{R}^{i+\delta}_z(\alpha_b) &\mbeq& \mathbf{R}^{i+\delta}_z(\beta) \ \ \forall \beta \in S'_\text{out} \label{eq:rin_rout2}.
\end{eqnarray}
resulting in
\begin{equation}
    \mathbf{R}_z^i(\alpha_b) + \delta t'(\beta) \mathbf{A}_z^i(\alpha_b) = \mathbf{R}_z^{i}(\beta) 
    + \delta t'(\beta) \mathbf{A}_z^i(\beta) \ \ \forall \beta \in S'_\text{out}
\end{equation}

At the boundary the collision between the temporally first in/out particle pair 
results in an exchange of the $z$-components of the propagating
vectors.
The exchange is 
\begin{eqnarray}
    \mathbf{a}_\perp^b = 
    &=& (\mathbf{a}_\text{out} - \mathbf{a}_\text{in})_z  \label{eq:collision1b} \\
    \mathbf{a}_{\text{out},z}' &=& \mathbf{a}_{\text{out},z} - \frac{2\ m_\text{in}}{m_\text{in} + m_\text{out}}\mathbf{a}_z\\
    \mathbf{a}_{\text{in},z}' &=& \mathbf{a}_{\text{in},z} + \frac{2\ m_\text{out}}{m_\text{in} +
    m_\text{out}}
    \mathbf{a}_z \label{eq:collision3b}
\end{eqnarray}
followed by an update of the propagating vector fields and the coordinate space in the same manner as described in the preceding sections.

No rotation operation is required in the case of an asymmetric surface model -- where solvent is in contact with the surface one side -- such as the one
studied in this work, see Section~\ref{s:lj_surface}. 
In the case of a symmetric surface model with solvent
particles in contact with the surface on both sides, 
$\mathbf{a}_\text{out}$ needs to be inverted 
if the conflicting in/out pair $\alpha_\text{b}$ and $\beta_\text{b}$ are located on opposite sides of the surface model.
$\mathbf{a}_\text{out}$ is then inverted once more after an
exchange of the propagating vectors.


\subsubsection{Multiatomic particles}

Applications require multiatomic particles to be handled as well. An example are the rigid TIPnP family of 
water force fields.\cite{abascal2005,jorgensen_comparison_1983,mahoney_five-site_2000}
In such cases SAFIRES works with reduced vector field spaces, corresponding to the position
and momentum at the center of mass for each molecular species. Effectively, each molecule 
is considered as a pseudo-particle and conflicts are monitored relative to the center of mass
for each pseudo-particle, not individual atoms. In the case of a system of $n_\text{mol}$ rigid water molecules the vector fields transform from $3n$- to $3n_\text{mol}$-dimensional fields, which are for the position and velocity  
\begin{align}
    \mathbf{R}^i_\text{cm} = \left\{ \sum_a^{S_\text{mol}}
    m_a \mathbf{R}^i(a)/M\ \forall S_\text{mol} \in S \right\}  \\
    \mathbf{V}^i_\text{cm} = \left\{ \sum_a^{S_\text{mol}}
    m_a \mathbf{V}^i(a)/M\ \forall S_\text{mol} \in S \right\}
\end{align}
where the subset $S_\text{mol}=\{\ce{O},\text{H1},\text{H2}\}$ indexes the atoms of the each molecule in the set $S$ of all atom indices. The force vector field is reduced as
\begin{equation}
    \mathbf{F}_\text{cm}(\mathbf{R}_\mathrm{cm}^i) =
    \left\{ \sum_a^{S_\text{mol}}
    \mathbf{F}(\mathbf{R}^i(a))\ \forall S_\text{mol} \in S  \right\}
\end{equation}
$m_a$ and $M$ is the mass of the atom and molecule, respectively.
Additionally, for the Langevin integrator the random variable vector fields transform as
\begin{equation}
    \mathbf{\Sigma}^i_\text{cm} = \left\{\sum_a^{S_\text{mol}}
    \sqrt{m_a} \mathbf{\Sigma}^i(a)/M\ \forall S_\text{mol} \in S  \right\} 
\end{equation}
and similarly for $\mathbf{\Lambda}^i_\text{cm}$.
The square root dependence of the per-atom mass-weighing follows from the
multiplicative factor $\bm{\sigma}$ as defined in equation \eqref{eq:sigma}. Using the center of mass transformed fields above, the corresponding propagating field, $\mathbf{A}^i_\text{cm}$, is evaluated with equation \eqref{eq:A}.

The SAFIRES processes operate on the reduced dimensional vector fields and corresponding molecular index set in such a way that the 'checking for conflicts', equations~\eqref{eq:boundary}-\eqref{eq:conflict}, followed by 'time step extrapolation', equations~\eqref{eq:solve_this}-\eqref{eq:dt_solution} result 
in identifying conflicting in/out pseudo-particle pair indices $\alpha_b$ and $\beta_b$, as well as index sets $S_\text{mol}^{\alpha_b} = \{\alpha^\text{O}_b,\alpha^\text{H1}_b,\alpha^\text{H2}_b\}$ and $S_\text{mol}^{\beta_b}=\{\beta^\text{O}_b,\beta^\text{H1}_b,\beta^\text{H2}_b\}$. 
That is, for a conflicting in/out pseudo-particle pair the molecular indices, and corresponding atomic indices are identified.


With the molecular index identified the SAFIRES 'elastic collision', as described in Section~\ref{s:collision},
results in an exchange of center of mass normal components.
For example, given conflicting psuedo-particle pairs and corresponding propagating fields
$\mathbf{a}^i_\text{in,cm}$ = $\mathbf{A}_\text{cm}^i(\alpha_b)$ and $\mathbf{a}^i_\text{out,cm} = \mathbf{A}_\text{cm}^i(\beta_b)$
the resulting center of mass components become
\begin{eqnarray}
    \mathbf{a}_\text{out,cm}'^{\text{\ rot}} &=& \mathbf{a}_\text{out,cm}^\text{rot} - \frac{2\ M_\text{in}}{M_\text{in} + M_\text{out}}\mathbf{a}_{\perp,\text{cm}}^b\\
    \mathbf{a}_\text{in,cm}' &=& \mathbf{a}_\text{in,cm} + \frac{2\ M_\text{out}}{M_\text{in} +
    M_\text{out}}
    \mathbf{a}_{\perp,\text{cm}}^b\label{eq:collision3}
\end{eqnarray}
After rotation of the outer pseudo-particle propagating vector -- to give $\mathbf{a}_\text{out,cm}'$ -- the components need to be redistributed to the atoms. The tangential components of the atoms in each pseudo-particle are not affected in the collision, hence the redistribution is done as follows
\begin{align}
    \mathbf{A}^i_c(\alpha) = \mathbf{a}_{\parallel,\text{in}}(\alpha) + \frac{m_\alpha}{M} (\mathbf{a}'_\text{in,cm})_\perp
    \ \forall\ \alpha \in S_\text{mol}^{\alpha_b} \\
    \mathbf{A}^i_c(\beta) = \mathbf{a}_{\parallel,\text{out}}(\beta) + \frac{m_\beta}{M} (\mathbf{a}'_\text{out,cm})_\perp
    \ \forall\ \beta \in S_\text{mol}^{\beta_b}
\end{align}
where $\mathbf{a}_\parallel$ are the unperturbed tangential components. 
This way any tangential modes 
of the molecule are not quenched in the process. While this procedure does conserve momentum, some energy in the 
rotational degrees of freedom is 
transformed to energy in the
translational degrees of freedom.

\section{Computational Details} \label{s:computational}

SAFIRES and FIRES are implemented within the Python-based framework of the Atomic Simulation Environment (ASE) available under the GNU Lesser General Public License\cite{larsen2017}. A force constant of $k_\text{FIRES} = 500$~kcal~A$^{-2}$ is used for calculations using FIRES.

Test calculations are performed using the LJ potential available in ASE with simulation parameters for argon ($\sigma = 3.4$~\AA, $\epsilon = 120\ k_\text{B}$, $\rho = 1.374$~g~cm$^{-3}$, $T = 94.4$~K)\cite{rahman1964}. For MD simulations, the Velocity-Verlet and Langevin dynamics implemented in ASE are utilized. The $g(r)$ of LJ systems are sampled over 1~ns, using an \textit{NVE} ensemble and a time step of 1~fs with the Velocity Verlet propagator. Starting configurations for LJ liquid simulations are pre-equilibrated over 1~ns of \textit{NVT} Langevin dynamics using the same time step. 

In case of the LJ liquid simulation, 25 of the total 512 LJ particles are included in the inner region, including the fixed central particle ("solute") for simulations using FIRES or SAFIRES. For the LJ surface model, three layers á 16 particles are cut in 111 direction from the most stable \textit{fcc} Ar crystal configuration\cite{barrett_crystal_1965}. The Ar surface model is frozen during simulations. The inner and outer regions contain 32 and 96 LJ particles in the liquid state, respectively.

The TIP4P force field is used for water simulations\cite{abascal2005}. The $g(r)$ for water-in-water calculations are sampled every 1~ps for a total of 20~ns, using an \textit{NVT} ensemble and the Langevin propagator with a time step of 0.5~fs. A friction coefficient of 0.05 is used. Starting configurations for water-in-water simulations are equilibrated over 250~ps in an \textit{NVT} ensemble, using the Langevin propagator with a time step of 2~fs and a friction coefficient of 0.01. The first 20 ps of each run are discarded before sampling. RATTLE constraints are used to ensure rigid bond lengths and bond angles of the water molecules\cite{andersen1983}. For calculations with FIRES and SAFIRES, 14 molecules are included in the inner region, including the fixed central molecule ("solute"). All pair distributions are sampled using the VMD program\cite{humphrey1996}.

\section{Test calculations} \label{s:test}

\subsection{Lennard-Jones liquid} \label{s:lj_liquid}

\begin{figure}[h!]
    \centering
    \includegraphics[width=0.5\linewidth]{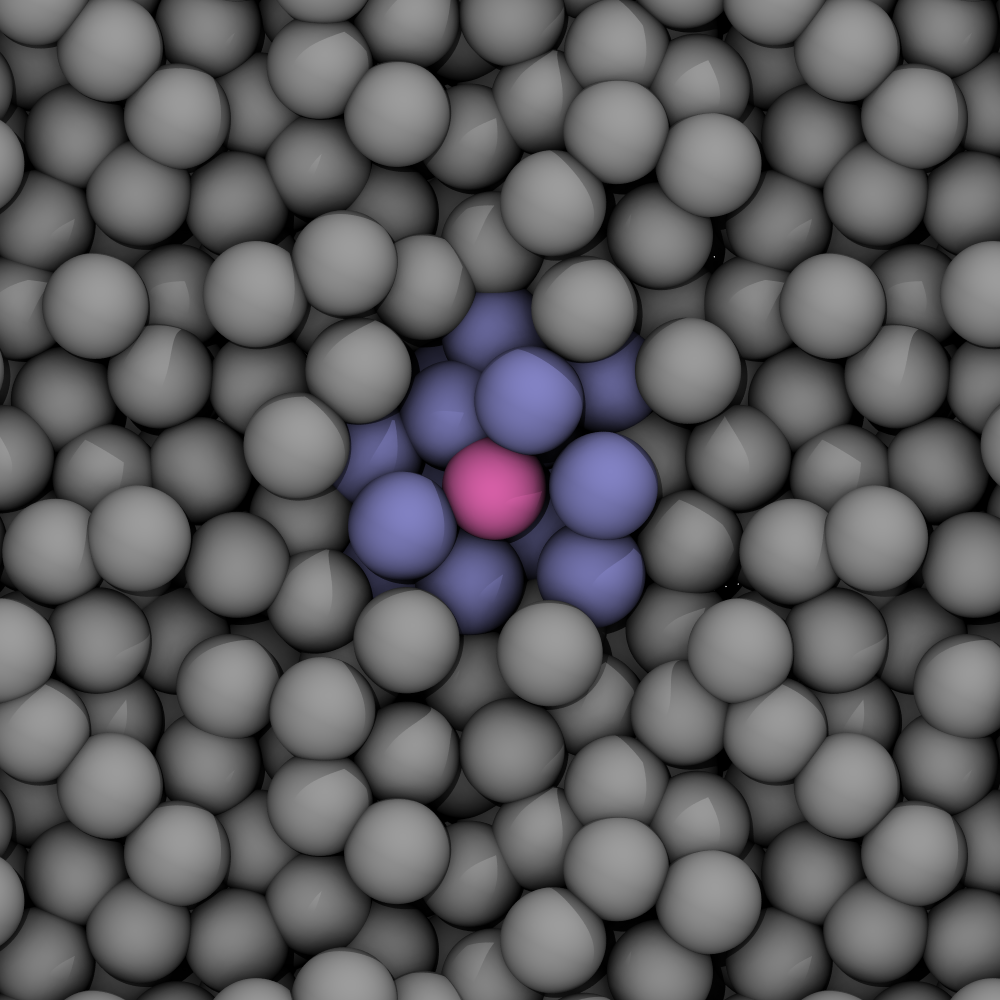}
    \caption{Illustration of a cross section of the LJ liquid computational model used in this work. Pink: solute, blue: particles in the inner region, grey: particles in the outer region.}
    \label{fig:lj_liquid_illustration}
\end{figure}

To test SAFIRES, an LJ liquid with argon parameters is used as the model system. This simple model allows efficient benchmarking of the technical implementation. Figure \ref{fig:lj_liquid_illustration} depicts a cross section of the liquid with different colors for the solute (pink), particles in the inner region (blue), and particles in the outer region (grey).
Both inner and outer region are calculated using the same LJ potential.

First, energy conservation of the SAFIRES algorithm is tested for \textit{NVE} Velocity-Verlet dynamics and this model system using the approach presented by Allen and Tildesley\cite{allen2017}. It is found that energy conservation with SAFIRES is significantly improved over FIRES, see Figure S1.

\begin{figure}[h!]
    \centering
    \includegraphics[width=0.7\linewidth]{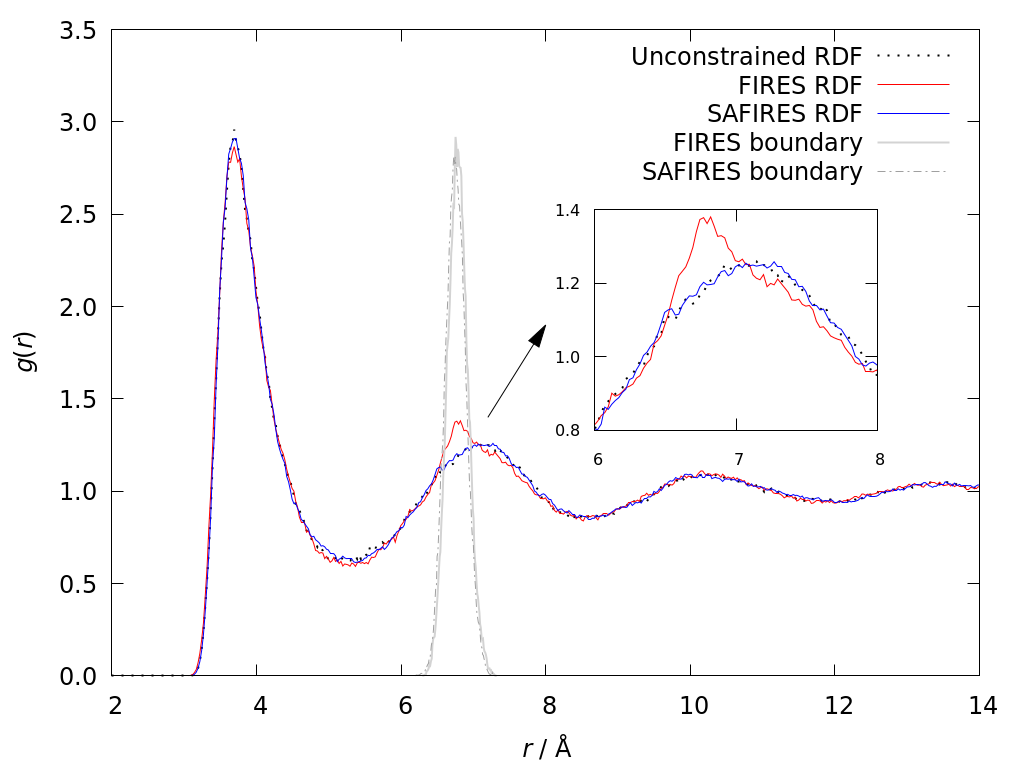}
    \caption{$g(r)$ of a LJ liquid using argon parameters. Pairs are sampled between a fixed central LJ particle ("solute") and the surrounding LJ particles. Black dotted line: reference calculation without any ensemble separation; red: FIRES; blue: SAFIRES; dashed light grey: normalized probability distribution of the SAFIRES boundary location; dark grey: normalized probability distribution of the FIRES boundary location.}
    \label{fig:lj_liquid_rdfs}
\end{figure}

Next, the influence of the ensemble separator on the simulation is quantified. To this end, the $g(r)$ are calculated from 1~ns of \textit{NVE} Velocity-Verlet dynamics each. Velocity-Verlet is used over Langevin to keep the first test as simple as possible and to monitor energy conservation. Only pairs involving the solute are considered so that $r$ correlates to the location of the boundary with respect to the solute. The $g(r)$ for an unconstrained reference simulation as well as for simulations with FIRES and SAFIRES are shown in Figure \ref{fig:lj_liquid_rdfs}.
Also shown in Figure \ref{fig:lj_liquid_rdfs} are normalized probability distributions of the locations of the FIRES and SAFIRES boundaries.

SAFIRES reproduces the unconstrained $g(r)$ without significant deviations. Notably, FIRES introduces accumulation artifacts around the boundary location. The boundary location distributions as well as the artifact around the boundary observed with FIRES broaden for a LJ liquid of lower density, see Figure S2. This means that the artifact can potentially be obfuscated under certain simulation conditions, as speculated by Bulo and co-workers\cite{bulo2013}. The observed improvement in case of SAFIRES is likely related to its instantaneous resolution of conflicts. When using FIRES, particles experience a spring force when passing the boundary, accelerating them either away or towards the solute. This means, they will spend several iterations in the other respective region, being first decelerated and then accelerated again in the opposite direction. It is suggested that the spring force approach will lead to an accumulation of density in the $g(r)$ around the border, even for a simplistic model system like the LJ fluid. Analogous results are obtained with an \textit{NVT} ensemble and Langevin dynamics (see Figure S3).

\subsection{Lennard-Jones surface} \label{s:lj_surface}

\begin{figure}[h!]
    \centering
    \includegraphics[width=0.5\linewidth]{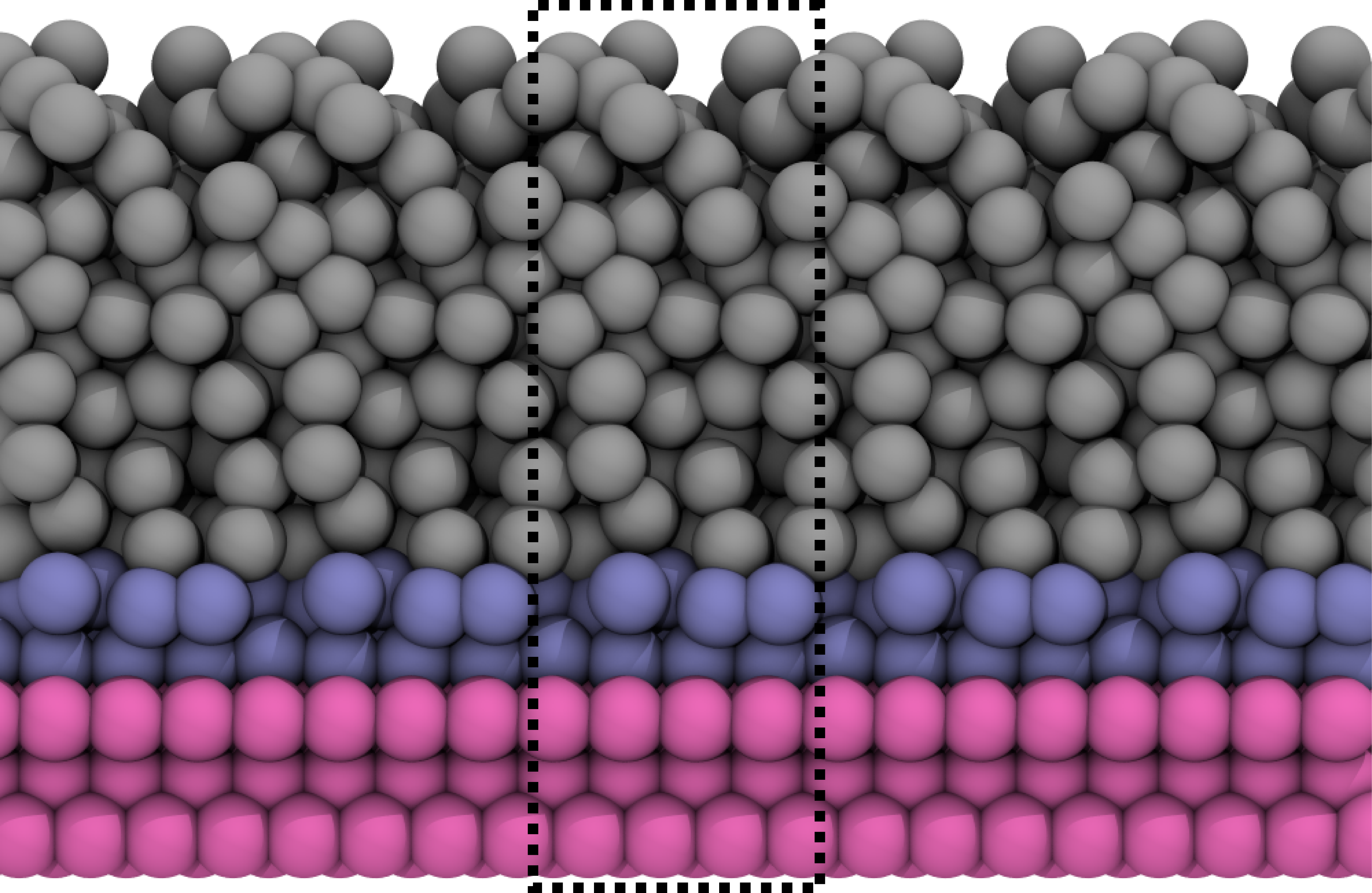}
    \caption{Illustration of the periodic surface computational model used in this work. Pink: frozen surface layer, blue: particles in the inner region, grey: particles in the outer region. Dotted line indicates the periodically repeating unit. Note that vacuum separates periodic images in \textit{z} direction.}
    \label{fig:lj_surface_illustration}
\end{figure}

SAFIRES is built with interface simulations in mind. To illustrate this application, a LJ liquid using argon parameters is placed in a simulation cell with a 111-indexed surface made up of three layers of frozen LJ argon particles. Figure \ref{fig:lj_surface_illustration} illustrates the model system.
The vacuum region above this asymmetric surface slab used to avoid interaction with the next periodic image in \textit{z} direction is not shown.

\begin{figure}[h!]
    \centering
    \includegraphics[width=0.7\linewidth]{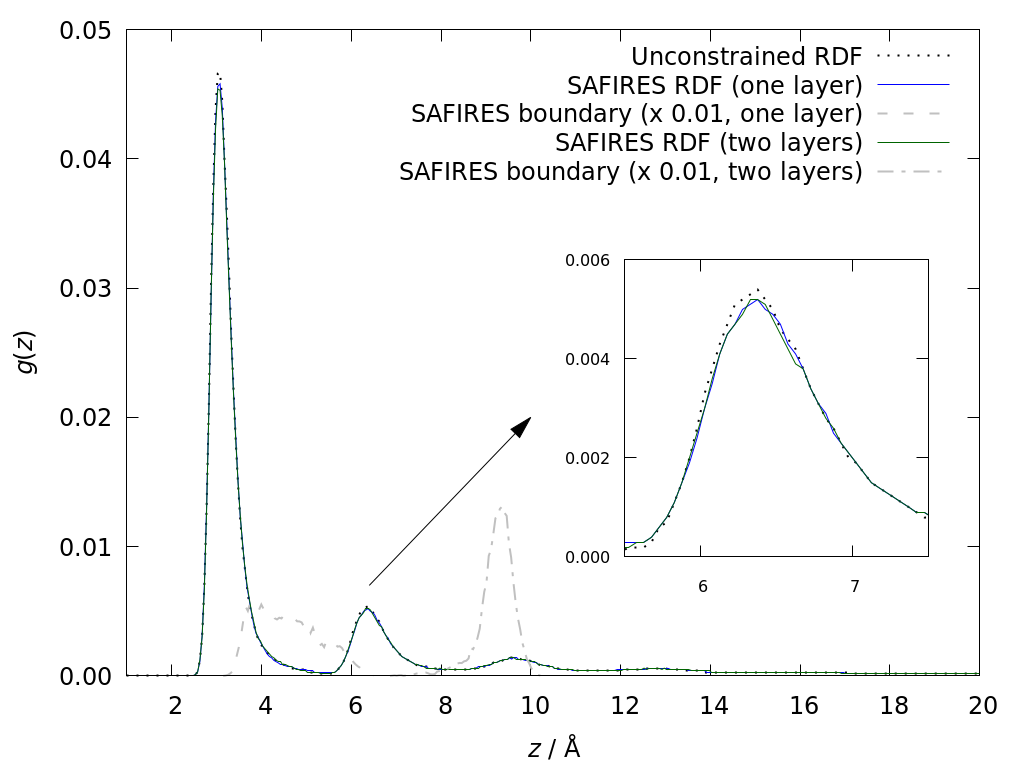}
    \caption{$g(z)$ of a LJ liquid using argon parameters on top of a three-layer (111)-indexed surface of frozen LJ argon particles. "One layer": 16 LJ liquid particles in the inner region; "two layers": 32 LJ particles in the inner region. Pairs are sampled between an \textit{xy} plane located in the top layer of the surface model and the LJ particles that constitute the liquid. Black dotted line: unconstrained reference calculation; blue: SAFIRES with 16 particles in the inner region; grey dashed line: normalized probability distribution of the SAFIRES boundary location corresponding to 16 particles in the inner region; grey dashed-dotted line: normalized probability distribution of the SAFIRES boundary location corresponding to 32 particles in the inner region.}
    \label{fig:lj_surface_rdfs}
\end{figure}

The $g(z)$ is calculated from 1~ns of \textit{NVE} Velocity-Verlet dynamics using this model. Since the solute in this case is a periodic surface, pairs are calculated between a \textit{xy} plane located in the top layer of the surface and the particles that constitute the liquid. This way, the origin of the abscissa in Figure \ref{fig:lj_surface_rdfs} coincides with the top layer of the surface model. Figure \ref{fig:lj_surface_rdfs} shows the resulting $g(z)$ of an unconstrained simulation and of two simulation using SAFIRES, one simulation with 16 LJ particles in the inner region ("one layer") and another simulation with 32 particles in the inner region ("two layers").
Simulations using SAFIRES reproduce the reference RDF without significant deviations, both in case of 16 or 32 particles within the inner region. This highlights the robustness of the method against the composition of inner and outer region.

In case of the simulation with 32 atoms in the inner region ("two layers"), the normalized probability distribution of the SAFIRES boundary is slightly broadened and asymmetric compared to the liquid presented in Figure \ref{fig:lj_liquid_rdfs}. This difference in shape is a result of the generally lower density of the liquid in this case as it can expand into the vacuum above the surface.

For the simulation with 16 atoms in the inner region, the normalized probability distribution of the SAFIRES boundary is significantly broadened as a result of interface effects. Analyzing the trajectories reveals that the first layer of LJ liquid particles ontop the surface behaves significantly more orderly than the remaining liquid, similar to the ice-like water layers on a Pt(111) surface in contact with liquid water. This leads to on average larger particle distances within this first layer and between the first and second layers and thus results in a broadening of the boundary location distribution in the \textit{z} direction perpendicular to the surface.

\subsection{Simulation of water-in-water} \label{s:tip4p}

\begin{figure}[h!]
    \centering
    \includegraphics[width=0.7\linewidth]{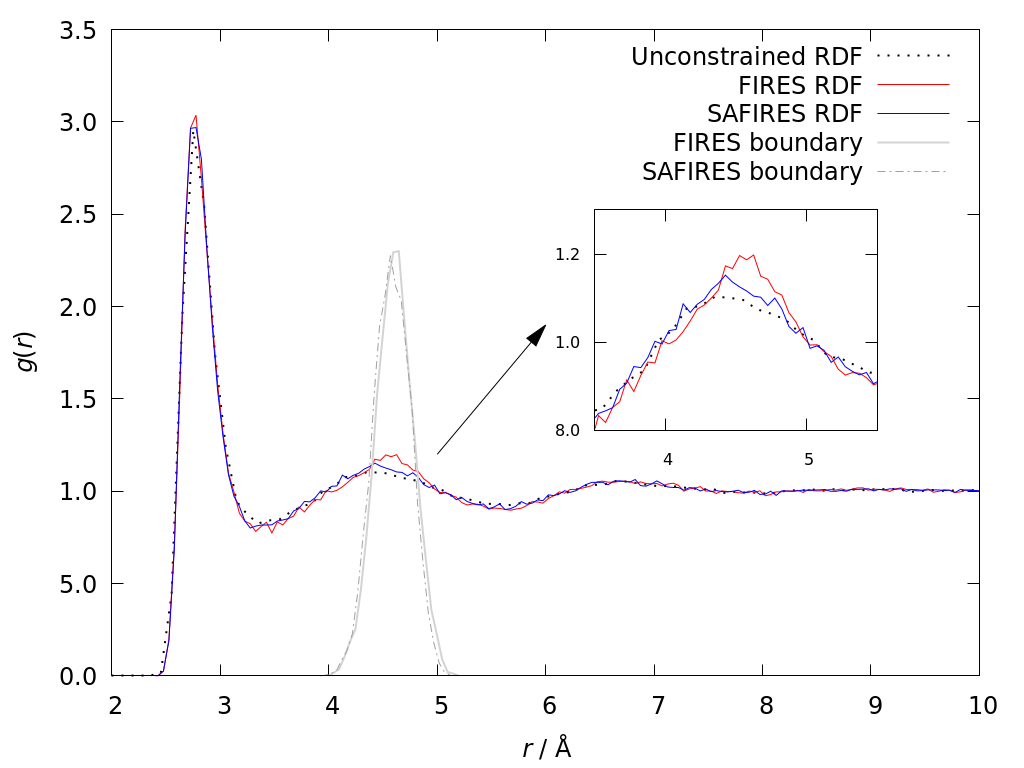}
    \caption{$g(r)$ of TIP4P water. Pairs are sampled between a fixed central water molecule ("solute") and the surrounding water molecules. Black dotted line: unconstrained reference calculation; red: FIRES; blue: SAFIRES; dashed light grey and dark grey: normalized probability distribution of the SAFIRES and FIRES boundary location, respectively.}
    \label{fig:tip4p_rdfs}
\end{figure}

Finally, exemplary MM*/MM simulations of water-in-water using the TIP4P force field and an \textit{NVT} ensemble are performed using FIRES and SAFIRES to separate the inner and outer regions. The geometry of water molecules is kept rigid as the present implementation of TIP4P does not support flexible molecules. The $g(r)$ of a reference calculation without ensemble separation as well as of simulations using FIRES and SAFIRES are shown in Figure \ref{fig:tip4p_rdfs}.
SAFIRES reproduces the unconstrained $g(r)$ without significant deviations. The $g(r)$ obtained with FIRES shows an accumulation artifact around the boundary region, similar to results obtained for the LJ liquid in Figure \ref{fig:lj_liquid_rdfs}.

\section{Conclusion}

The scattering-adapted flexible inner region ensemble separator (SAFIRES) is a separation algorithm for hybrid simulations coupling different methodologies. It has been designed in particular for interface calculations. Like other boundary-based methods, SAFIRES is built on the premise that if the same type of particle is present in the inner and outer region, correct average ensemble statistics can be obtained despite particles being unable to cross the boundary and exchange between regions. However, unlike other boundary-based methods that use repulsive forces or bias potentials to keep particles separated, SAFIRES instantaneously redirects particles at the boundary via energy- and momentum-conserving elastic collisions. This approach therefore constitutes the most rigorous implementation of the original flexible boundary premise by Rowley and Roux to date\cite{rowley2012}. To ensure exact collisions, SAFIRES introduces a new, multiple-time-step propagator which reduces to the Vanden-Eijnden / Ciccotti implementation of a Langevin propagator for constant time steps and to the Velocity-Verlet algorithm for constant time steps and zero friction.

Using a LJ liquid and a LJ liquid in contact with a surface as exemplary systems, SAFIRES reproduces unconstrained reference $g(r)$ or $g(z)$, respectively, without significant deviations while FIRES introduces accumulation artifacts around the boundary region. Lastly, an exemplary MM*/MM study of water-in-water using the TIP4P force field is presented. The $g(r)$ obtained with SAFIRES does not significantly deviate from the unconstrained reference calculation while accumulation artifacts are present in a simulation using FIRES. SAFIRES will be available as part of the open-source ASE package.

Future development will focus on the inclusion of virtual counter ions in SAFIRES to model the electrode potential near a surface within a Poisson-Boltzmann scheme as well as combining SAFIRES with polarizable QM/MM coupling models\cite{dohn2019} for electrochemical applications. SAFIRES is also not restricted to liquid-liquid or solid-liquid interfaces; it could for example be used to study diffusion in solids for battery research. Future work will therefore also focus on exploring various interface phenomena using this approach.

\begin{acknowledgement}

BK thanks the University of Iceland Research Fund for funding through a PhD fellowship. Computations were performed on resources provided by the Icelandic High Performance Computing Centre at the University of Iceland. This work was supported by the Icelandic Research Fund and by the DFG (German Science Foundation) through the Collaborative Research Center SFB-1316 as well as the cluster of excellence POLiS (project ID 390874152).

\end{acknowledgement}

\begin{suppinfo}

The results of the energy conservation test, a RDF of a Lennard-Jones liquid at lower density, a RDF of a Lennard-Jones liquid sampled using an NVT ensemble, and pseudocode of the SAFIRES algorithm are presented in the electronic Supporting Information.

\end{suppinfo}

\bibliography{bibliography}

\end{document}




\newpage

\section{Energy conservation test}

\begin{figure}[h!]
    \centering
    \includegraphics[width=0.7\linewidth]{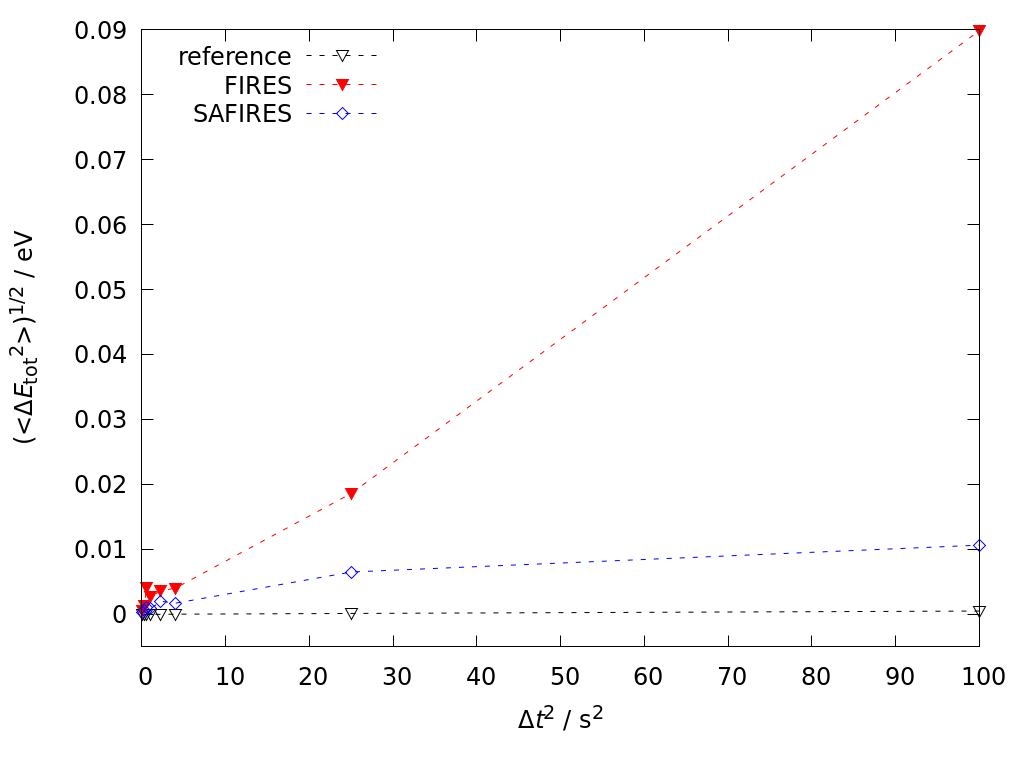}
    \caption{Energy conservation test for unconstrained reference simulations (black empty triangles), simulations using the FIRES boundary method (red filled triangles) and simulations using the SAFIRES boundary method (blue empty diamonds) on the basis of the LJ liquid model system presented in Figure 3.}
    \label{fig:conservation}
\end{figure}

Energy conservation is tested with the LJ liquid model system ($\sigma = 3.4$~\AA, $\epsilon = 120\ k_\text{B}$, $\rho = 1.374$~g~cm$^{-3}$, $T = 94.4$~K) shown in Figure 3, using the Verlocity-Verlet propagator (\textit{NVE} ensemble) and time steps between 0.25 and 10~fs for  a total of 100~ps according to the scheme proposed by Allen and Tildesley\cite{allen2017}. Results for each time step are averaged over 5 runs from different starting images. The obtained energy conservation trends are shown in Figure \ref{fig:conservation}.
The calculated RMS energy fluctuation trends indicate that energy conservation is improved using SAFIRES over FIRES.

\section{Lennard-Jones Liquid at lower density}
To illustrate the effect of varying density on the position and width of the FIRES and SAFIRES boundary location probability distribution, the simulation illustrated in Figure 3 of the main manuscript is repeated using a lower density of the LJ liquid ($\rho = 1.100$~g~cm$^{-3}$ \textit{vs.} $\rho = 1.374$~g~cm$^{-3}$). The resulting $g(r)$ for a FIRES simulation, a SAFIRES simulation, and an unconstrained reference simulation are shown in Supplementary Figure \ref{fgr:lj_lowdens}.
\begin{figure}[h!]
    \centering
    \includegraphics[width=0.7\linewidth]{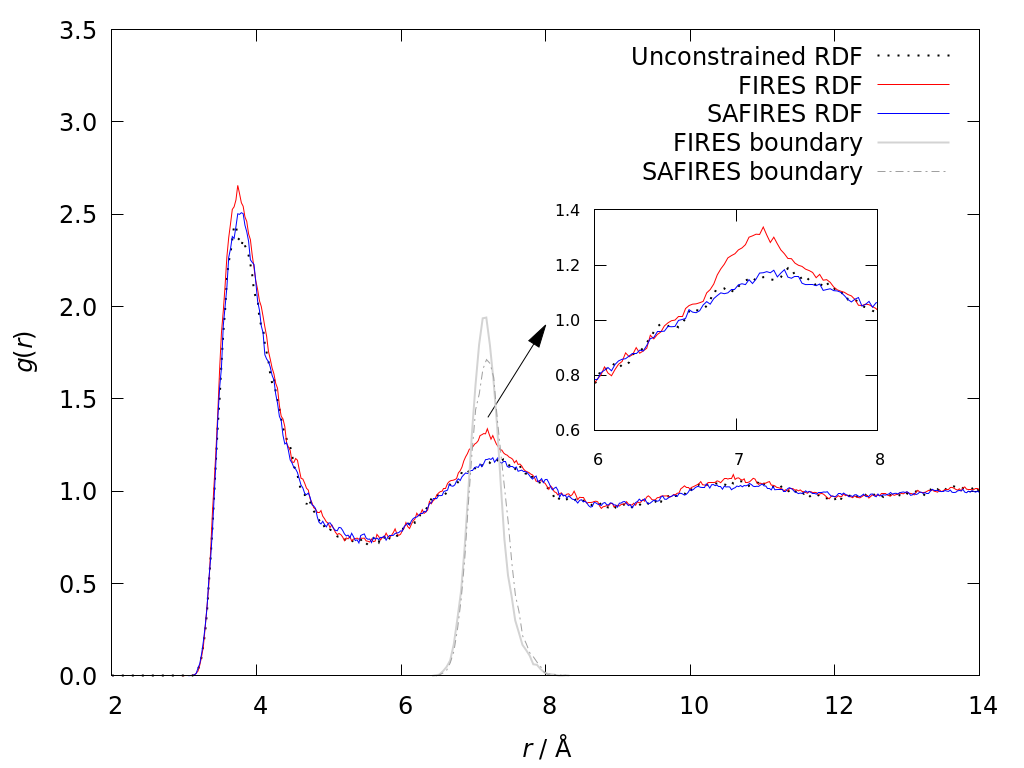}
    \caption{$g(r)$ of a LJ liquid using argon parameters. Black dashed line: reference simulation without any ensemble separation; red: simulation using FIRES boundary method; blue: simulation using SAFIRES boundary method; dashed light grey: probability distribution of the FIRES boundary location; dark grey: probability distribution of the SAFIRES boundary location. This example uses a lower density of the liquid ($\rho = 1.100$~g~cm$^{-3}$) than the example in the main manuscript.}
    \label{fgr:lj_lowdens}
\end{figure}
$g(r)$ are sampled over 1~ns using a timestep of 1~fs and the argon LJ parameters outlined in the computational section of the main manuscript.

\section{Lennard-Jones Liquid, \textit{NVT} ensemble}
\begin{figure}[h!]
    \centering
    \includegraphics[width=0.7\linewidth]{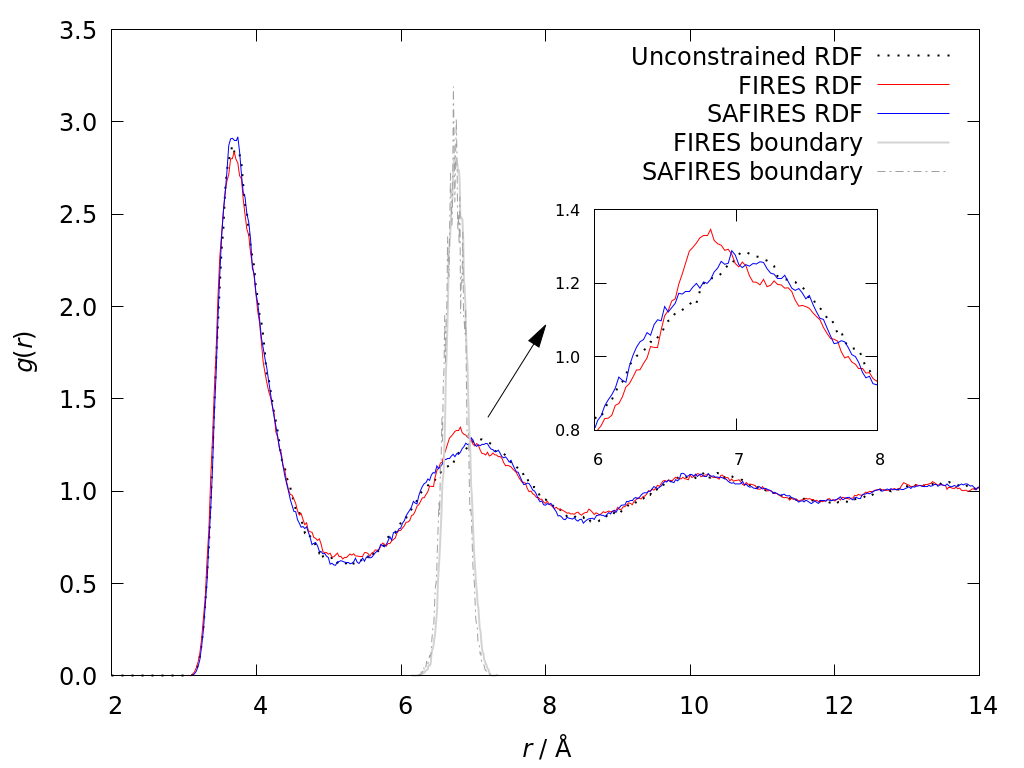}
    \caption{$g(r)$ of a LJ liquid using argon parameters. \textit{NVT} ensemble. Black dashed line: reference simulation without any ensemble separation; red: simulation using FIRES boundary method; blue: simulation using SAFIRES boundary method; dashed light grey: probability distribution of the FIRES boundary location; dark grey: probability distribution of the SAFIRES boundary location.}
    \label{fgr:nvt}
\end{figure}

The LJ liquid model system depicted in Figure 3 ($\sigma = 3.4$~\AA, $\epsilon = 120\ k_\text{B}$, $\rho = 1.374$~g~cm$^{-3}$) is simulated using an \textit{NVT} ensemble and the Langevin propagator. A friction coefficient of 0.05 is used to achieve the thermostat target temperature of $94.4$~K. Figure \ref{fgr:nvt} displays the resulting $g(r)$ of an unconstrained reference simulation as well as for simulations using FIRES and SAFIRES.


\clearpage

\section{SAFIRES pseudocode}

$atoms$: MD output array containing properties and calculation results for all atoms.\\
$previous\_atoms$: $atoms$ array from a previous iteration.\\
$\Delta t$: default time step of the MD simulation.\\
$\Delta t_\text{rem}$: float used to keep track of fractional time steps.\\
$recent$:  array used to remember the particle pair that recently collided.\\
\vspace{0.5cm}
Langevin equation as implemented in ASE:
\begin{eqnarray}
    v' &=& v + \frac{1}{2}\ \Delta t\ F - \frac{1}{2}\ \Delta t\ \gamma\ v + \frac{1}{2}\ \sqrt{\Delta t}\ \sigma\ \xi \nonumber \\
    &-& \frac{1}{8}\ \Delta t^2\ \gamma \left( F - \gamma\ v \right) - \frac{1}{4}\ \Delta t^\frac{3}{2}\ \gamma\ \sigma \left( \frac{1}{2}\ \gamma + \frac{1}{\sqrt{3}}\ \eta \right)\\
    x' &=& x + \Delta t\ v' + \Delta t^\frac{3}{2}\ \sigma\ \frac{1}{2 \sqrt{3}}\ \eta\\
    \text{Where:   } \sigma &=& \sqrt{\frac{2\ T\ \gamma}{m}}
\end{eqnarray}
$x$: array of position vectors for each atom.\\
    $m$: array of masses for each atom.\\
    $v$: array of velocity vectors for each atom.\\
    $F$: array of force vectors for each atom.\\
    $T$: simulation temperature.\\
    $\gamma$: Langevin friction coefficient.\\
    $\xi$, $\eta$: array of random force vectors for atom $x$.\\

\begin{algorithm}[H]
\scriptsize
\SetAlgoLined
\KwResult{Find conflicts (outer atom inside inner area) and coordinate their resolution.}
 \vspace{0.5cm}
 \For{$inner$ atom in $atoms$}{
    calculate distance $d_{inner}$ to solute\;
 }
 set $d_\text{thresh}$ = largest calculated $d_{inner}$ value\;
 set $i_{\text{thresh}}$ = index of inner atom corresponding to largest $d_{inner}$\;
 \vspace{0.5 cm}
 
 $conflicts$ = array()\;
 \For{outer atom in $atoms$}{
   $d_{outer}$ = distance of outer atom to solute\;
    \If{$d_{outer} \leq d_\text{thresh}$}{
        $i_\text{outer}$ = index of outer atom\;
        \If{($i_\text{outer}$, $i_\text{thresh}$) != $recent$}{
            $\Delta t_\text{new}$ = \textbf{extrapolate}($i_\text{outer}$, $i_\text{thresh}$)\;
            add array($i_{outer}$, $idx_{thresh}$, $\Delta t_\text{new}$) to $conflicts$\;
        }
    }
 }
 \vspace{0.5 cm}
 \eIf{length($conflicts$) $>$ 0}{
    choose array($i_{outer}$, $i_{thresh}$, $\Delta t_\text{new}$) from $conflicts$ that contains the smallest $\Delta t_\text{new}$ value\;
    $atoms$ = $previous\_atoms$\;
    \textbf{propagate}($atoms$, $\Delta t_\text{new}$, $halfstep=1$, $recheck$)\;
    \textbf{collide}($i_\text{outer}$, $i_\text{thresh}$)\;
    $recent$ = array($i_\text{outer}$, $i_\text{thresh}$)\;
    $previous\_atoms = atoms$\;
    \eIf{$\Delta t_\text{rem}$ == 0}{
        $\Delta t_\text{rem}$ = $\Delta t - \Delta t_\text{new}$\;
    }{
        $\Delta t_\text{rem} = \Delta t_\text{rem} - \Delta t_\text{new}$\;
    }
    \textbf{propagate}($atoms$, $\Delta t_\text{rem}$, $halfstep=2$, $recheck$)\;
    \textbf{check}($atoms$, $recheck$=True)\;
 }{
    \newpage
    $previous\_atoms$ = atoms\;
    $\Delta t_\text{rem}$ = 0\;
    $recent$ = empty\;
 }
 
 \caption{check($atoms$, $recheck$=False)}
\end{algorithm}

\newpage

\begin{algorithm}[H]
\scriptsize
\SetAlgoLined
\KwResult{Extrapolate the timestep required to propagate the conflicting inner and outer particles to the same distance wrt. the solute.}
    \vspace{0.5cm}
    Update $x$, $v$, $F$, $\xi$, $\eta$, $m$ from $atoms$\;
    \vspace{0.5cm}
    \If{$recheck$ == False}{
        \For{$x$ in array($i_\text{outer}$, $i_\text{thresh}$)}{
            $\sigma_x = \sqrt{2\ T\ \gamma\ /\ m_x}$\;
            $e_x = \Delta t\ (F_x - \gamma\ v_x)\ /\ 2 + \sqrt{\Delta t}\ \sigma_x\ \xi_x\ /\ 2$\\
            $\ \ \ \ \ \ \ \ \ - \Delta t^2\ \gamma\ (F_x - \gamma\ v_x)\ /\ 8 
                               - \Delta t^\frac{3}{2}\ \gamma\ \sigma_x\ (\xi_x\ /\ 2 + \eta_x\ /\ \sqrt{3})\ /\ 4$\;
            $f_x = \Delta t^{\frac{1}{2}}\ \sigma_x\ \eta_x\ /\ (2 \sqrt{3})$\;
        }
    }
    \If{$recheck$==True}{
        $e = 0$\;
        $f = 0$\;
    }
    $v_{outer}' = v_{i_{outer}} = v_{outer} + e_{outer} + f_{outer}$\;
    $v_{inner}' = v_{i_{thresh}} = v_{inner} + f_{inner} + e_{inner}$\;
    \vspace{0.5cm}
    $c_0 = \left < r_{outer}, r_{outer} \right > - \left < r_{inner}, r_{inner} \right > $\;
    $c_1 = 2 \left < r_{outer}, v_{outer}' \right > - 2 \left < r_{inner}, v_{inner}' \right > $\;
    $c_2 = \left < v_{outer}', v_{outer}' \right > - \left < v_{inner}', v_{inner}' \right > $\;
    \vspace{0.5cm}
    $results$ = array()\;
    \For{root in $(c_2\ x^2 + c_1\ x + c_0 = 0)$}{
        \If{real($root$) == $True$ AND $root > 0$ AND $root <= \Delta t$}{
            save $root$ in $results$\;
        }
    }
    \textbf{return} smallest value in $results$ as $\Delta t_\text{new}$\;
\caption{extrapolate($i_\text{outer}$, $i_\text{thresh}$, $recheck$)}
\end{algorithm}

\newpage

\begin{algorithm}[H]
\scriptsize
\SetAlgoLined
\KwResult{Propagate atoms using a Langevin style propagator. Both halfsteps can be propagated separately.}
    \vspace{0.5cm}
    Update $x$, $v$, $F$, $\xi$, $\eta$, $m$ from $atoms$\;
    \vspace{0.5cm}
    \If{$recheck$ == False}{
        $\sigma = \sqrt{2\ T\ \gamma\ /\ m}$\;
        $e = \Delta t\ (F - \gamma\ v)\ /\ 2 + \sqrt{\Delta t}\ \sigma\ \xi\ /\ 2$\\
        $\ \ \ \ \ \ \ \ \ - \Delta t^2\ \gamma\ (F - \gamma\ v)\ /\ 8 
                       - \Delta t^\frac{3}{2}\ \gamma\ \sigma\ (\xi\ /\ 2 + \eta\ /\ \sqrt{3})\ /\ 4$\;
        $f = \Delta t^\frac{1}{2}\ \sigma\ \eta\ /\ (2 \sqrt{3})$\;
    }
    \If{$recheck$ == True}{
        $e = 0$\;
        $f = 0$\;
    }
    \vspace{0.5cm}
    \If{$halfstep == 1$}{
        $v' = v + e$\;
        $x' = x + \Delta t_{new}\ (v' + f)$\;
        overwrite atomic positions in $atoms$ with $x'$\;
        $v'' =  (x' - x - \Delta t_{new}\ f) / \Delta t_{new}$\;
    }
    \If{$halfstep == 2$}{
        $x' = x + \Delta t_{new}\ v$\;
        overwrite atomic positions in $atoms$ with $x'$\;
        $v' =  (x' - x) / \Delta t_{new}$\;
        calculate new forces $\rightarrow$ $F = F_\text{new}$\;
        re-calculate $e$ with $F_\text{new}$ (see above)\;
        $v'' = v' + e$\;
    }
    update atomic velocities in $atoms$ with $v''$\;
    
\caption{propagate($atoms$, $\Delta t_\text{new}$, $halfstep$, $recheck$)}
\end{algorithm}

\newpage

\begin{algorithm}[H]
\scriptsize
\SetAlgoLined
\KwResult{Perform a fully elastic collision between the conflicting inner and a outer atoms.}
    \vspace{0.5cm}
    $r_{outer}, r_{inner}$: distance vector from solute to $inner$, $outer$.\\
    $m_{outer}, m_{inner}$: atomic mass of $inner$, $outer$.\\
    $v_{outer}, v_{inner}$: velocity vector of $inner$, $outer$.\\
    \vspace{0.5cm}
    $\theta = arctan2 \left ( \frac{r_{outer} \times r_{inner}}{\left | r_{outer} \times r_{inner} \right |}, \left < r_{outer}, r_{inner} \right > \right )$\;
    $n = \frac{r_{outer} \times r_{inner}}{\left | r_{outer} \times r_{inner} \right |}$\;
    $a = \cos(\theta)$\;
    $b, c, d = -n\ \sin(\theta)$\;
    $X^\text{rot} = 
     \begin{bmatrix}
        a^2 + b^2 - c^2 - d^2 & 2 (bc + ad) & 2 (bd - ac) \\
        2 (bc - ad) & a^2 + c^2 - b^2 - d^2 & 2 (cd + ab) \\
        2 (bd + ac) & 2 (cd - ab) & a^2 + d^2 - b^2 - c^2
    \end{bmatrix}$\;
    $r_{outer}^\text{rot} = \left < X^\text{rot}(\theta), r_{outer} \right > $\;
    $v_{outer}^\text{rot} = \left < X^\text{rot}(\theta), v_{outer} \right > $\;
    \vspace{0.5cm}
    $M = m_{outer} + m_{inner}$\;
    $v_{12} = v_{outer}^\text{rot} - v_{inner}$\;
    $v_{outer}^\text{rot, post} = v_{outer}^\text{rot} 
            - \frac{2 m_{inner}}{M}\ \frac{\left < v_{12}, r_{inner} \right > \cdot r_{inner}}{|r_{inner}|^2}$\;
    $v_{inner}^\text{post} = v_{inner} 
            + \frac{2 m_{outer}}{M}\ \frac{\left < v_{inner}, r_{inner} \right > \cdot r_{inner}}{|r_{inner}|^2}$\;
    $v_{outer}^\text{post} = \left < X^\text{rot}(2 \pi - \theta), v_{outer}^\text{rot, post} \right > $\;
    apply new velocities to $atoms$\;
    
\caption{collide($idx_\text{outer}$, $idx_\text{inner}$)}
\end{algorithm}

\clearpage
\bibliography{supplementary-bibliography}